\documentclass[
 superscriptaddress,
preprint,
 amsmath,amssymb,
 aps,
prb,
]{revtex4-2}

\usepackage{graphicx}
\usepackage{dcolumn}
\usepackage{bm}
\usepackage{hyperref}
\begin{document}
\title{Advancing AI-Driven Analysis in X-ray Absorption Spectroscopy: Spectral Domain Mapping and Universal Models}
\author{Nina Cao}
\thanks{These authors contributed equally to this work.}
\affiliation{Department of Mechanical Engineering, Massachusetts Institute of Technology, 77 Massachusetts Ave, Cambridge, MA 02139, USA}

\author{Pavan Ravindra}
\thanks{These authors contributed equally to this work.}
\affiliation{Department of Chemistry, Columbia University, 3000 Broadway, New York, New York 10027, USA}

\author{Shubha R. Kharel}
\affiliation{Computing and Data Sciences Directorate, Brookhaven National Laboratory, Upton, New York 11973, USA}

\author{Chuntian Cao}
\affiliation{Computing and Data Sciences Directorate, Brookhaven National Laboratory, Upton, New York 11973, USA}

\author{Boyang Li}
\affiliation{Computing and Data Sciences Directorate, Brookhaven National Laboratory, Upton, New York 11973, USA}

\author{Xuance Jiang}
\affiliation{Center for Functional Nanomaterials, Brookhaven National Laboratory, Upton, New York 11973, USA}
\affiliation{Department of Physics and Astronomy, Stony Brook University, Stony Brook, NY 11794, USA}

\author{Matthew R. Carbone}
\affiliation{Computing and Data Sciences Directorate, Brookhaven National Laboratory, Upton, New York 11973, USA}

\author{Xiaohui Qu}
\affiliation{Center for Functional Nanomaterials, Brookhaven National Laboratory, Upton, New York 11973, USA}

\author{Deyu Lu}
\email{dlu@bnl.gov}
\affiliation{Center for Functional Nanomaterials, Brookhaven National Laboratory, Upton, New York 11973, USA}

\date{\today}

\begin{abstract}
In recent years, rapid progress has been made in developing artificial intelligence (AI) and machine learning (ML) methods for x-ray absorption spectroscopy (XAS) analysis. Compared to traditional XAS analysis methods, AI/ML approaches offer dramatic improvements in efficiency and help eliminate human bias. To advance this field, we advocate an AI-driven XAS analysis pipeline that features several inter-connected key building blocks: benchmarks, workflows, databases, and AI/ML models. Specifically, we present two case studies for XAS ML. In the first study, we demonstrate the importance of reconciling the discrepancies between simulation and experiment using spectral domain mapping (SDM). Our ML model, which is trained solely on simulated spectra, predicts an incorrect oxidation state trend for Ti atoms in a combinatorial zinc titanate film. After transforming the experimental spectra into a simulation-like representation using SDM, the same model successfully recovers the correct oxidation state trend. In the second study, we explore the development of universal XAS ML models that are trained on the entire periodic table, which enables them to leverage common trends across elements. Looking ahead, we envision that an AI-driven pipeline can unlock the potential of real-time XAS analysis to accelerate scientific discovery.

\end{abstract}

\newcommand{\mat}[1]{\mathbf{#1}}
\maketitle

\section{Introduction}
X-ray absorption spectroscopy (XAS) is a widely used materials characterization technique across a broad range of domain sciences, including physics, chemistry, material science, and biology~\cite{rehr2000theoretical,penner1999x,yano2009x}. In XAS measurements, core electrons are excited by incident x-ray photons to the empty energy states of the sample. Within the multiple scattering picture, XAS describes how the photoelectron created at the absorbing atom is scattered by the neighboring atoms~\cite{penner1999x}. As such, XAS is a powerful element-specific probe of the local chemical environment at the atomic scale, which encodes information about the absorbing atom (e.g. the strength and phase shift of the scattering potential) as well as its surroundings along different scattering paths. 

However, quantitative analysis of XAS spectra, specifically the x-ray absorption near edge structure (XANES), remains technically challenging due to the complex correlation between local atomic structures and their corresponding XAS features. Empirical approaches, like the empirical fingerprint method~\cite{gaur2015speciation}, are generally limited by the available experimental reference spectra. Computational XAS can provide detailed analysis of spectral shapes and perform peak assignment~\cite{rehr2000theoretical,de2008core}, but it requires extensive domain knowledge and, in the case of first-principles methods, is time consuming and demands a large amount of computational resources.

Recent progress has been made in AI-driven XAS analysis to address these limitations, leveraging the rapid development of AI/ML tools. The first demonstration of a successful ML analysis was made by Timoshenko \emph{et al.}, who trained a supervised ML model on simulated Pt L-edge XANES data to predict average coordination numbers up to the fourth shell of Pt nanoparticles~\cite{timoshenko2017supervised}. The ML model was trained on simulation and generalized successfully to the experimental data, as it was able to predict the shape and size of the Pt nanoparticles from XAS measurements. Since then, the field of XAS analysis has witnessed rapid adoption of ML methods in various applications, including analyzing spectral fingerprints~\cite{aarva2019understanding,aarva2019understanding2,jiang2025resolving}, predicting XAS spectra from atomic structure (the forward problem)~\cite{carbone2020machine,rankine2020deep,rankine2022accurate,ghose2023uncertainty,kotobi2023integrating,kwon2023harnessing,kharel2025omnixas,GleasonCuXASNet}, and extracting physical descriptors from XAS spectra (the inverse problem)~\cite{marcella2020neural,liu2021probing,liu2019mapping,guda2021understanding,chen2024robust,routh2021latent,xiang2022solving,trejo2019elucidating,tetef2021unsupervised, carbone2019classification,torrisi2020random,gleason2024prediction}. Generative methods, such as variational / adversarial autoencoders~\cite{tetef2021unsupervised,liang2023decoding} and diffusion models~\cite{kwon2024spectroscopy}, have also been adopted to enhance the sampling of the spectral latent space or the atomic configuration space. The success of these studies highlights the power of AI/ML in resolving the complex structure-spectrum relationship in XAS.

\begin{figure*}[tbh!]
    \centering
    \includegraphics[width=2.5 in]{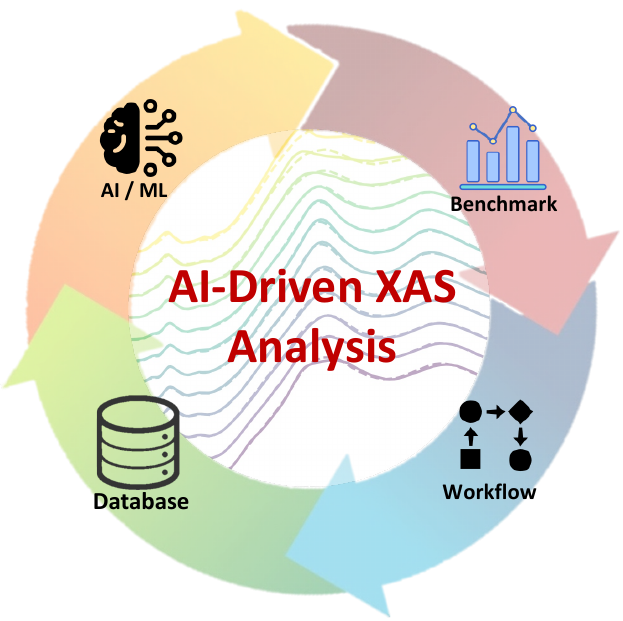}
    \caption{Schematic of the AI-driven XAS analysis pipeline.}
    \label{fig:ai-framework}
\end{figure*}

To streamline and foster more broad applications of AI/ML in XAS analysis, further efforts must be made to develop the infrastructure of the pipeline. Figure~\ref{fig:ai-framework} highlights several key building blocks that are essential to this paradigm. 

\begin{itemize}

\item \textbf{Benchmarks:} Systematic benchmarks are needed to quantify the accuracy of theoretical predictions and identify areas for improvement. The first step is to benchmark among different theoretical methods and computer codes, which by itself is non-trivial as it requires collaborative efforts from multiple research groups. So far, only a few multi-code benchmark studies have been carried out, e.g., for 3d transition metal L-edge~\cite{de20212p} and Ti K-edge XAS~\cite{meng2024multicode}. Such efforts need to be extended to broader material spaces. More importantly, systematic benchmarks between theory and experiment are still lacking.

\item \textbf{Workflow Software:} High-throughput spectral simulations enable efficient sampling of large materials spaces with well-defined atomic structure labels, making them the method of choice for constructing training sets for XAS ML models. Know-hows learned from the Benchmark module, such as input parameters to ensure numerical convergence of simulations, can be incorporated into the simulation workflow to lower the entry barrier for non-experts. Workflow software, such as \emph{Lightshow}~\cite{carbone2023lightshow} and \emph{Corvus}~\cite{kas2021advanced}, helps to streamline generation of input files for XAS simulation and ensure data reproducibility and interoperability.  

\item \textbf{Databases:} Data availability and quality are essential in ML applications. Several public XAS databases generated with the multiple scattering code FEFF~\cite{rehr2010parameter} have been developed for materials and molecules, such as the Materials Project~\cite{mathew2018high,chen2021database} and the QM9 database~\cite{ghose2023uncertainty}. More XAS databases have become available for specific material families, such as 3d transition metal K-edge~\cite{torrisi2020random,kharel2025omnixas,rankine2022accurate}, Fe complexes K-edge~\cite{guda2021understanding} and S K-edge in  lithium thiophosphate solid electrolyte~\cite{guo2023simulated}. However, existing XAS databases do not yet adequately span the entire space of interest for molecules and materials, and XAS database development remains an active research direction. Although large FEFF spectral databases have been the testbed for ML models, it is necessary to also develop databases with other methods, such as the core-hole potential method~\cite{karsai2018effects,taillefumier2002x}, time-dependent density functional theory~\cite{neese2012orca,valiev2010nwchem}, and the Bethe-Salpeter equation~\cite{vinson2011bethe,gulans2014exciting}. For practical applications on experimental data, simulated spectra need to be carefully validated against experimental standards.

\item \textbf{ML Models:} Existing XAS ML models fall mainly into two categories. The first is ``proof-of-principles" models, which use databases of simulated spectra for both the forward and inverse problems ~\cite{carbone2020machine,rankine2020deep,rankine2022accurate,ghose2023uncertainty,kotobi2023integrating,kharel2025omnixas,GleasonCuXASNet,tetef2021unsupervised, kwon2023harnessing,carbone2019classification,torrisi2020random}. Typically, FEFF is used to generate spectra because the low computational cost makes it the first choice for building large spectral databases. The second category involves ML models that are trained on simulation data but applied to experimental spectra. Typically, these models consider a narrow chemical space~\cite{timoshenko2017supervised,marcella2020neural,liu2021probing,liu2019mapping,guda2021understanding,chen2024robust,routh2021latent,xiang2022solving,trejo2019elucidating, gleason2024prediction}. In these cases, the corresponding spectra have few degrees of freedom, which helps reduce the complexity of the problem. Despite these efforts, developing general-purpose ML models that are applicable to experimental data in broad molecular and materials spaces remains an open research question.
\end{itemize}

The development of an AI-driven XAS analysis pipeline (Benchmarks, Workflow Software, Databases, and ML Models) follows an iterative procedure. When a new type of material is encountered that does not belong to the previous training set, first-principles simulations are carried out and the results are validated against experiment. New simulated spectral data are used to augment the spectral training set and XAS ML models are retrained. The diagram in Figure~\ref{fig:ai-framework} goes in a loop, making the pipeline adaptive to new structures and spectral shapes.

To enhance the fidelity and broad applicability of AI-driven XAS analysis, we list several important future research directions.
\begin{enumerate}
    \item Systematic benchmarking of XAS simulation against experiment that quantifies the accuracy of the employed approximations for different physical effects, such as core-hole final-state effects, multiplet effects, finite-temperature effects, and satellite effects.
    \item Assembling large-scale XAS databases that combine both accurate simulations and experimental standards.
    \item Developing methods to address the remaining discrepancies between simulation and experiment in order to enhance the transferability of ML models.
    \item Training foundation XAS models on literature and curated XAS data to provide domain scientist-level expertise for XAS analysis.
\end{enumerate}
In this perspective, we highlight items 3 and 4 with two examples. Our first example is associated with the idea of bridging the gap between simulation and experiment, and the second example explores the idea of universal XAS models for the whole periodic table.

\section{Bridging XAS Experiment and Simulation in Machine Learning}

A major challenge in applying supervised ML to XAS analysis is the scarcity of labeled experimental spectra. To overcome this limitation, most existing ML models are instead trained on simulated spectra. This raises the question of how to reconcile the intrinsic discrepancies between experimental and simulated XAS data. In this section, we highlight the need to bridge these two data domains and propose a simple method for aligning experimental and simulated spectra. While we focus on Ti K-edge XANES spectra, our approach readily generalizes to multi-element datasets.

\subsection{Experimental vs. Simulated Spectra}

There are several fundamental differences between experimental and simulated XAS spectra. Experimental spectra measure absorption as an average over the various local environments of a given absorbing element. Although this can provide rich information about materials with heterogeneous local environments, we often lack precise knowledge of the site-specific atomic structures that needs to be deconvoluted from the average spectra. This makes it difficult to extract site-specific structural labels like coordination number (CN) and oxidation state (OS) from experimental spectra. Because supervised ML models require these labels at training, experimental spectra alone are generally not well suited for supervised ML. Moreover, assembling large crowd-sourced databases of experimental spectra is challenging, since measurements from different groups often vary due to differences in instrumentation, sample preparation methods, and measurement modes (e.g., transmission vs. fluorescence).

Simulated spectra, on the other hand, are obtained for specific atomic structures by calculating a band structure or solving a model Hamiltonian. In either case, the local chemical environment is well defined, which gives us direct access to precise, site-resolved structural labels. These calculations can also be standardized using high-throughput approaches, which has enabled the construction of large public datasets of simulated XAS spectra~\cite{mathew2018high,chen2021database,ghose2023uncertainty}. The size and label-rich nature of these datasets make them ideal for training supervised ML models. A major caveat, however, is the discrepancy between simulation and experiment, which depends strongly on the level of approximations used to treat relevant XAS effects, such as core-hole final state effects, multiplet effects, finite temperature effects, and shake-up satellites.

\begin{figure*}[tbh!]
    \centering
    \includegraphics[width=3.375in]{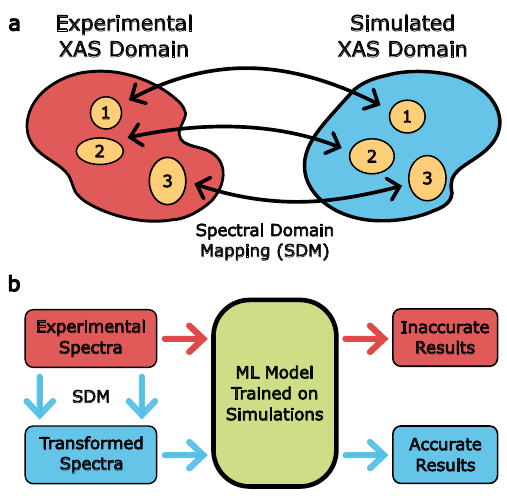}
    \caption{Schematic of spectral domain mapping (SDM). (a) SDM learns a transformation between corresponding regions in the experimental and simulated XAS domains. Sub-regions correspond to different chemical environments. (b) Workflow of ML applications with and without the SDM.}
    \label{fig:pavan_overview}
\end{figure*}

While simulated data are ideal for training AI models, our goal is to deploy these models to analyze the experimental spectra of real materials. The distinction between these two spectral domains presents a core technical challenge for AI-driven XAS analysis. Because experimental and simulated spectra follow different distributions from a data science perspective, models trained on simulations may produce unreliable predictions when applied directly to experimental data. To address this modality gap, we present a data-driven spectral domain mapping (SDM) approach, outlined in Figure~\ref{fig:pavan_overview}. The SDM transforms experimental spectra into a simulation-like representation, which improves the compatibility of the input spectra with the ML model's training distribution. We show that the SDM enables more accurate downstream predictions of local chemical descriptors from XAS spectra.

\subsection{Material System: Combinatorial Zinc Titanate Thin Film}

\begin{figure*}[h!]
    \centering
    \includegraphics[width=3.375in]{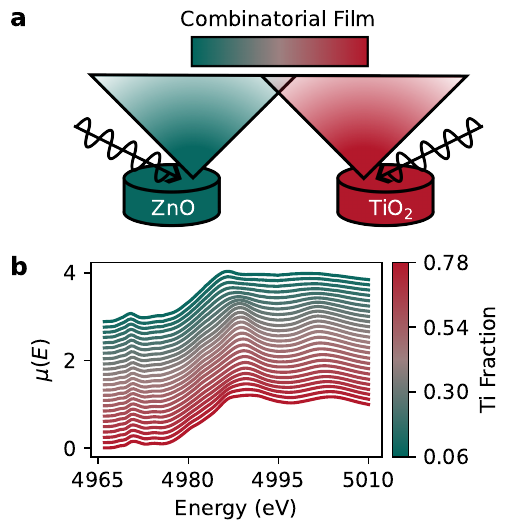}
    \caption{(a) The combinatorial thin film of zinc titanate synthesized with pulsed laser deposition. (b) Sample Ti K-edge XANES spectra measured along the Ti concentration gradient.}
    \label{fig:pavan_deposition}
\end{figure*}
 
As a case study, we apply ML-based analysis techniques to a thin film of combinatorial zinc titanate. Combinatorial materials are of great practical interest because their physical properties can be tuned continuously by varying the relative concentrations of the underlying metal constituents. In this work, we consider a combinatorial zinc titanate thin film prepared using pulsed laser deposition (PLD). The PLD technique can create smoothly varying composition profiles in the synthesis of single-wafer samples~\cite{li2023deciphering}.

A detailed description of the PLD process can be found in Ref.~\citenum{li2023deciphering}. Here we provide a brief summary of the procedure, as shown in Figure~\ref{fig:pavan_deposition}. First, a high-energy pulsed laser vaporizes a target metal oxide into small particles that deposit on one side of a substrate wafer. The concentration of the metal oxide tapers off as the distance from the vaporization site increases, which naturally creates a concentration gradient. Subsequently, the wafer is rotated 180 degrees and another pulsed laser vaporizes the second metal oxide onto the wafer. By alternating the vaporization and deposition of the two metal oxides, we obtain a thin film on the wafer with a spatial composition gradient. For the zinc titanate film considered in this work, this results in a gradual variation of the Zn:Ti ratio across the wafer's surface. Figure~\ref{fig:pavan_deposition}c shows a sequence of 25 down-sampled experimental Ti K-edge XANES spectra measured at different positions along the wafer, labeled by their local Ti fraction~\cite{li2023deciphering}.

The resulting thin film exhibits a rich variety of zinc titanate domains with diverse local atomic environments. Practically, this allows for a rapid and thorough screening of material properties across a wide range of Zn:Ti ratios, all on a single wafer. From a methodology perspective, this combinatorial sample serves as an excellent testbed for developing robust XAS analysis methods that are applicable across a diverse range of local structures~\cite{jiang2025resolving}.

\subsection{ML Model: Rank-Constrained Adversarial Autoencoder}

\begin{figure*}[h!]
    \centering
    \includegraphics[width=\textwidth]{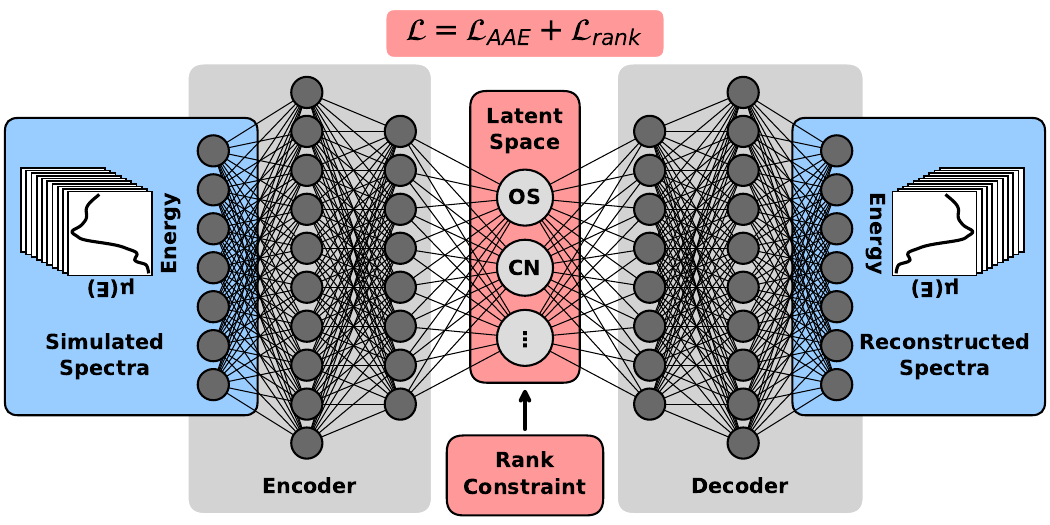}
    \caption{The RankAAE autoencoder architecture. The RankAAE loss function $\mathcal{L}$ combines a standard adversarial autoencoder (AAE) loss with a rank constraint that aligns each latent dimension with a chosen chemical descriptor, such as oxidation state (OS) or coordination number (CN).}
    \label{fig:pavan_RankAAE}
\end{figure*}

To predict the OS and CN of the Ti atoms in the zinc titanate film, we train a rank-constrained adversarial autoencoder (RankAAE) model~\cite{liang2023decoding}. RankAAE is a semi-supervised autoencoder that compresses XAS spectra into a low-dimensional continuous latent space, as depicted in Figure~\ref{fig:pavan_RankAAE}. In addition to the typical reconstruction loss, RankAAE's loss function includes a second term that aligns each latent dimension with a chosen structural descriptor, such as OS or CN. Because the loss function requires these site-specific labels, our training dataset consists of simulated Ti K-edge XANES spectra and their associated OS and CN labels~\cite{kharel2025omnixas}. These simulated spectra were generated using the core-hole potential method~\cite{karsai2018effects} implemented in the Vienna Ab initio Simulation Package (VASP)~\cite{kresse1993ab}.

The latent space of the RankAAE model has six dimensions. As a result of the rank constraint, five of these latent variables are strongly correlated with their corresponding physical descriptors (OS, CN, the average CN of the nearest-neighbor oxygen atoms, the nearest-neighbor radial standard deviation, and the minimum oxygen-oxygen distance of the nearest neighbors)~\cite{liang2023decoding}. The remaining latent variable is not included in the rank constraint, allowing the model to use it purely to aid in reconstructing the input spectra. In this study, we train an ensemble of 32 fully connected neural networks to map the latent space (six-dimensional vectors) to the OS and CN of structures in the training set. We verified that the model generalized well to the held-out validation set. The outputs of the model are real numbers, which should be interpreted as the averaged OS and CN of the local motifs in the material. The reported OS and CN values are given by the mean of the ensemble's predictions, with the standard deviation serving as an estimate of the error.

\begin{figure*}[h!]
    \centering
    \includegraphics[width=3.375in]{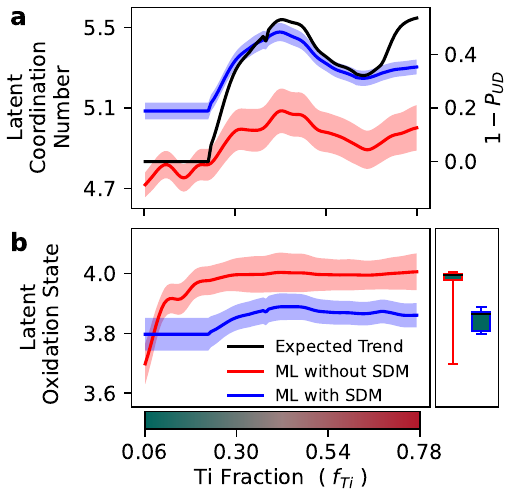}
    \caption{(a) Predicted CN from RankAAE without and with SDM. The expected trend, based on the proportion of under-coordinated or distorted Ti atoms, is shown as $1-P_{UD}$. (b) Predicted OS without and with SDM; side panel shows box plots of distribution of these OS values. The expected trend is that the OS should stay constant as Ti fraction varies. Shaded areas indicate the standard deviation across the ensemble of neural networks.}
    \label{fig:pavan_trends}
\end{figure*}

Figure~\ref{fig:pavan_trends} shows RankAAE's CN and OS predictions (red curves) when applied directly to the experimental Ti K-edge XANES spectra from the zinc titanate film. As shown in Figure~\ref{fig:pavan_trends}a, the expected CN trend is given by $1 - P_{UD}$ (black curve) based on our previous work~\cite{jiang2025resolving}, where $P_{UD}$ is the proportion of Ti species with under-coordinated (CN less than six) or distorted first shells. First-principles calculations indicate that the shape of $1 - P_{UD}$ is governed by the complex local motif evolution as the Ti concentration ($f_{Ti}$) varies~\cite{li2023deciphering,jiang2025resolving}. This evolution involves defect and amorphous regions with under-coordinated or distorted Ti, as well as crystalline domains such as TiZn$_2$O$_4$, where Ti forms octahedral motifs. As Figure~\ref{fig:pavan_trends}a shows, $1 - P_{UD}$ remains very low for $f_{Ti}< 20\%$, as the local structures in this concentration range are dominated by Ti defects in wurtzite ZnO. As $f_{Ti}$ increases further, TiZn$_2$O$_4$ domains become more prominent, driving CN upwards until $f_{Ti} \approx 39\%$. After that,  $1 - P_{UD}$ starts to decay gradually until $f_{Ti} \approx 71\%$ and increases again as rutile TiO$_2$ domains start to form. Importantly, $1 - P_{UD}$ remains below 0.5 across all Ti concentrations, as the zinc titanate film contains a significant amount of amorphous structure. The trend in $1 - P_{UD}$ is very well reproduced by RankAAE's CN predictions, including the low values for  $f_{Ti}< 20\%$, the peak at $f_{Ti} \approx 39\%$, and the relatively high values in the post-peak region.

However, the OS trend predicted by RankAAE is incorrect. Because the zinc titanate synthesis was performed under oxygen-rich conditions, Ti should be fully oxidized to Ti$^{4+}$ and remain constant across the wafer~\cite{li2023deciphering}. Instead, RankAAE predicts that the average OS of the Ti motifs increases from 3.7 to 4.0 in regions with low Ti fraction, leading to a broad and physically incorrect distribution of OS values across $f_{Ti}$. This suggests that the RankAAE model does not generalize well to experimental spectra of samples from regions with low Ti fraction, even though its predictions in high Ti fraction regions are consistent with experimental expectations. This failure likely originates from the discrepancy between simulated and experimental Ti K-edge XAS spectra, as the agreement between simulated VASP spectra and experiment is not perfect. Importantly, such breakdowns are difficult o anticipate beforehand, highlighting the need for a spectral transformation that ensures input spectra belong to the same distribution as the simulated spectra in the training dataset.

\subsection{Spectral Domain Mapping}
In Ref.~\citenum{jiang2025resolving}, we developed a cluster blind-signal-separation (cBSS) method to decompose experimental spectra using basis functions. The initial guess for the spectral basis functions is chosen to be the centroids of clusters of simulated spectra. These basis functions are then optimized with the BSS method to align with a small dataset of down-sampled experimental spectra. In this way, the optimized basis functions serve as an effective experimental spectral basis. The BSS method thus provides a convenient mapping between the simulated and experimental domains. We leverage this correspondence to propose an SDM method that maps experimental Ti spectra to simulation-like spectra so that they align better with the RankAAE training data.

The cBSS procedure represents each site-averaged spectrum as a linear combination of the spectra of representative local structures, i.e., spectral basis functions. We define a matrix $\mat{T}=\{\mu_1, \mu_2,\cdots,\mu_N\}$, where each column $\mu_i$ is an experimental spectrum, and a matrix $\mat{S}=\{s_1, s_2,\cdots,s_n\}$ of $n\le N$ spectral basis functions $s_j$. All spectra are represented as vectors of absorption coefficients on a common energy grid with $m$ points. We assume that each experimental spectrum in $\mat{T}$ is a weighted average of the underlying basis functions in $\mat{S}$. These non-negative weights are stored in a matrix $\mat{C}_{n\times N}$ whose columns sum to 1,

\begin{equation}
    \mat{T} = \mat{S} \, \mat{C}.
    \label{eqn:cBSS}
\end{equation}
For a given $\mat{C}$ matrix, the BSS procedure approximates $\mat{S}$ using the Moore-Penrose pseudoinverse of $\mat{C}$. The elements of $\mat{C}$ are optimized to minimize the Euclidean distance between this $\mat{S}$ matrix and an initial guess $\mat{\tilde{S}}=\{\tilde{s}_1, \tilde{s}_2,\cdots,\tilde{s}_n\}$ in the loss function,

\begin{equation}
    \mathcal{L} = \| \mat{\tilde{S}} - \mat{T} \mat{C}^{\dagger} \|,
\end{equation}
where ${}^{\dagger}$ indicates the pseudoinverse, given by $\mat{C}^{\dagger}=(\mat{C}^T \mat{C})^{-1} \mat{C}^T$.
In our previous work~\cite{jiang2025resolving}, the initial guess functions $\tilde{s}_j$ were chosen to be the centroids of clusters of site-specific simulated spectra. In this way, $\tilde{s}_j$ represent simulated spectral basis functions. The cBSS fitting procedure yields both the basis functions $\mat{S}$ and the weight matrix $\mat{C}$. 

In the SDM approach, the key idea is to transform the experimental spectra in $\mat{T}$ to simulation-like spectra $\mat{\tilde{T}}$ using the simulated spectral basis in $\mat{\tilde{S}}$, while retaining the same weights $\mat{C}$,
\begin{equation}
    \mat{\tilde{T}} = \mat{\tilde{S}} \, \mat{C}.
\end{equation}
In this study, $\mat{T}$ contains 100 experimental spectra of the zinc titanate combinatorial film shown in Figure~\ref{fig:pavan_deposition}b (i.e., $N=100$). Four representative spectral basis functions ($n=4$) were previously found to fully capture the spectral variation in the combinatorial zinc titante film, with their structural motifs characterized by first- and second-shell structures:  $\text{TiO}_2$,  $\text{Ti}_{\text{6H}}$,  $\text{Ti}_{\text{6L}}$, and $\text{Ti}_{\text{UD}}$~\cite{jiang2025resolving}. The $\text{TiO}_2$ motif represents the reference $\text{TiO}_2$ compound as a mixture of rutile and anatase; the $\text{Ti}_{\text{UD}}$ motif represents Ti species with under-coordinated or distorted first-shells; $\text{Ti}_{\text{6H}}$ and $\text{Ti}_{\text{6L}}$ motifs correspond to octahedral Ti species with high and low Ti fractions in the second shell, respectively. To account for differences due to the initialization of the parameters in $\mat{C}$, we average our results over 100 random initializations in the BSS procedure. The results vary minimally between different trials, suggesting that our fitting procedure is robust to the initialization of the $\mat{C}$ matrix.

\subsection{Results}

\begin{figure*}[h!]
    \centering
    \includegraphics[width=\textwidth]{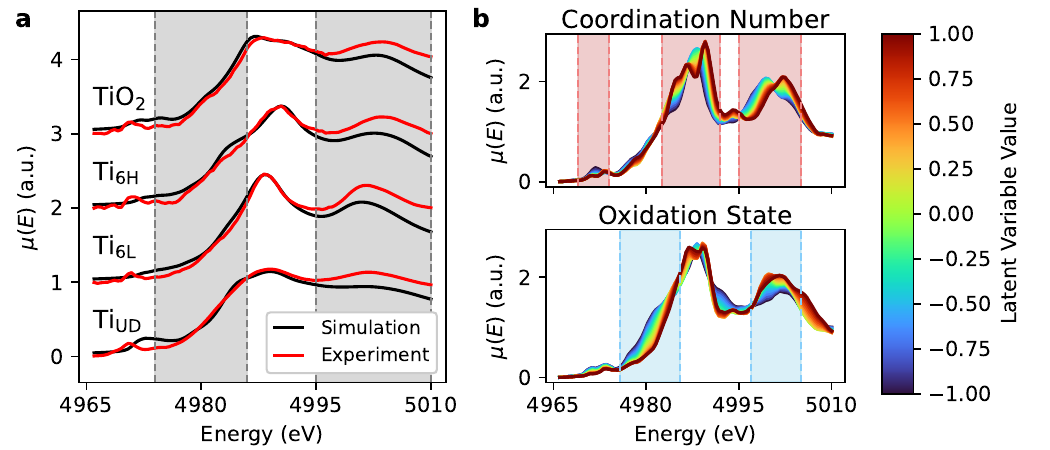}
    \caption{(a) Comparison of the simulated and optimized spectral basis functions. (b) The spectral variation with coordination number and oxidation state, as predicted by RankAAE. Highlighted regions in both panels emphasize regions with markedly different behavior.}
    \label{fig:pavan_latent}
\end{figure*}

In Figure~\ref{fig:pavan_latent}a, we compare the four simulated basis functions $\tilde{s}_j$ (black curves) with their corresponding optimized experimental basis functions $s_j$ (red curves) to understand the discrepancies between simulation and experiment. The shaded regions highlight the main differences: a shift in the edge position (particularly for $\text{Ti}_\text{6H}$ and $\text{Ti}_\text{6L}$) and a change in the height of the second peak around 5005 eV.

Since the RankAAE latent variables are aligned with structural descriptors, we can investigate how these spectral differences might influence downstream RankAAE predictions. Specifically, we can scan across values of a given latent variable to explore which spectral features RankAAE associates with the corresponding structural descriptor~\cite{li2023deciphering}. Figure~\ref{fig:pavan_latent}b shows that CN primarily affects the pre-edge peak, the height of the first peak, and the position of the second peak. OS, on the other hand, affects the edge position and the height of the second peak. Thus, the differences in the simulated and experimental basis spectra shown in Figure~\ref{fig:pavan_latent}a are in the same region as the spectral features that are associated with OS. Consequently, we might expect \textit{a priori} that the SDM procedure will have a larger impact on OS predictions than CN predictions.

The predicted CN and OS from using the SDM-transformed spectra as input are also shown in Figure~\ref{fig:pavan_trends}. The predicted CN trends are similar with or without SDM, and both show qualitative agreement with the expected trend (e.g. the shape of the curve) obtained in Ref.~\citenum{jiang2025resolving}. However, SDM qualitatively changes RankAAE's OS prediction. Without the SDM, RankAAE predicts that OS decreases sharply in the low $f_{Ti}$ region, and the resulting broad distribution of OS contradicts the fact that Ti sites should have a constant OS under the oxygen-rich synthesis conditions. In contrast, the OS predicted from the SDM-transformed spectra remains nearly constant across Ti fraction, in much better agreement with the expected trend. The small fluctuations observed are smaller than the associated error bars. The breakdown of RankAAE at low Ti fraction is likely driven by the shift in the absorption edge towards lower energies, as shown in the experimental spectra from Figure~\ref{fig:pavan_deposition}. The latent space analysis in Figure~\ref{fig:pavan_latent}b shows that such downshift in the edge position corresponds to lower OS values. RankAAE likely attributes this shift in the experimental absorption edge as evidence of a significantly lower OS, as the ML model may be oversensitive to this spectral feature. The simulated basis functions shown in Figure~\ref{fig:pavan_latent}a display much less variation in the onset of the edge, which helps explain why they yield a more uniform OS prediction.

This result underscores the utility of the SDM transformation in improving model fidelity on out-of-distribution experimental XAS spectra. It is worth nothing that the mean value of the predicted OS is around $+3.85$, which is slightly smaller than the nominal charge state of $+4$. Further improvement of the model may require using a continuous measure of the charge state, such as Bader charges instead of nominal charges, as the label~\cite{bader1985atoms} for OS.

\section{Learning Global Trends in XAS}
Traditionally, the analysis of experimental XAS data is performed for individual elements and edges, as the edge positions of different elements are often well-separated in energy. Consequently, existing XAS analysis methods target a specific element and edge in a given chemical/material space (e.g., Ti K-edge XAS of titanium oxides, as demonstrated in the previous section).

While these siloed XAS analysis methods work reasonably well in practice, an intriguing question is whether there is a benefit to develop universal models for XAS that are pre-trained on data from the whole periodic table. Such universal models can be readily applied to a wide range of materials and can be fine-tuned with system-specific data to further improve accuracy for target systems of interest. The concept of universal models for materials science has been demonstrated in areas such as neural network interatomic potentials like M3GNET \cite{chen2022universal}, and MACE-MP \cite{batatia2023foundation}. These models are trained on first-principles calculations of large material structure databases. They offer tremendous utility for material discovery in complex search spaces characterized by vast chemical diversity and large system size.
\subsection{Universal Model of XAS Spectra}
Since XAS is an abstract representation of the local chemical environment, smooth variation of the neighboring shell structures should be reflected as smooth changes in spectral features. XAS spectra also exhibit well established features beyond the element-wise core electron binding energies. For example, the edge position is a direct measure of the OS, known as Kunzl's law~\cite{kunzl1932linear}, i.e., a deeper edge position corresponds to a higher OS and vice versa. In 3d transition metal K-edge XANES, the pre-edge is derived from $s \rightarrow d$ dipole-forbidden transitions. The intensity of the pre-edge peak is sensitive to inversion symmetry. In systems with broken inversion symmetry, the mixing of $p$ and $d$ orbitals can give rise to stronger pre-edge peaks than for structures with inversion symmetry. In general, late 3d transition metals have weaker pre-edge peaks than early 3d transition metals because of the reduced amount of empty $d$ states. From established domain knowledge, one would expect to observe global trends in XAS (e.g. OS and local symmetry) that cross the boundaries between elements.

In our previous work on 3d transition metal K-edge XANES~\cite{kharel2025omnixas}, the uniform manifold approximation and projection (UMAP) method was used to construct a latent representation of the simulated FEFF spectral database~\cite{rehr2010parameter} with reduced dimensionality. Instead of performing independent analyses for each element, eight spectral families (Ti -- Cu) were analyzed together. We found that the spectral UMAP pattern was well separated by element type, oxidation state (OS), coordination number (CN), and oxygen coordination number (OCN). More importantly, the UMAP analysis revealed clear evidence of universality from the spectral latent space. Distinct sub-regions of physical descriptors within each elemental cluster often coalesced into larger clusters across elements. In fact, we found that a certain subset of spectra of one element could be more similar to spectra in different elemental clusters than to those within its own cluster\cite{kharel2025omnixas}. In the subsequent spectral prediction task using graph neural network models (known as OmniXAS), we found that the best performance is achieved by fine-tuning a universal model pre-trained on eight elements. The fine-tuned universal models improved prediction accuracy by up to 11\% over models trained on the target element alone~\cite{kharel2025omnixas}. 

\subsection{Data Curation and Cleaning}
Given the success of the universal model for 3d transition metals, we are motivated to extend this approach to the entire periodic table, or at the very least, a substantial portion of it, to effectively capture universal trends across elements in the spectral latent space.
As an initial step, we pulled site-specific FEFF spectra from hydrogen to antimony from the Materials Project K-edge XAS database \cite{mathew2018high} and analyzed the trends in the spectral latent space using UMAP. This raw database has 612,765 spectra.
Multiple post-processing steps were employed prior to the analysis in order to ensure data consistency and integrity, which resulted in a final set of 295,832 site spectra used for our analysis.
\begin{enumerate}
    \item Energy grid standardization: We standardized the energy grids by interpolating all spectra onto a uniform grid with a 35 eV range. For each element, the first energy grid point was chosen as the minimum energy across all the raw FEFF spectra. We used 200 grid points, which yields an energy resolution of 0.175 eV.
    \item Spectral selection: The raw FEFF database contains a mixture of normalized and unnormalized spectra, and the metadata lacks the information necessary to back-calculate the normalization constants. In order to avoid any potential data inconsistencies, we chose to use only normalized spectra, which comprises the majority of the database (approximately 76.4\%).
    \item Element filtering: We visually inspected heatmaps for all spectra corresponding to each element. Elemental heatmaps for elements heavier than Sb do not show discernible spectral line shapes. 
    In addition, the noble gas family is chemically inert and has too few spectra (He: 3, Ne: 1, Ar: 2 and Kr: 22), whereas most of other elements have thousands of spectra. For these reasons, we only retain the spectra of 47 elements (atomic numbers below 52 excluding noble gases).
    \item Duplicate removal: 
    To reduce sampling bias, exact duplicates of the spectra were removed across the board (approximately 4.8\% of the database or 29,420 spectra). Additionally, nearly duplicate spectra in Cd, Zn, Cr and S were removed when the spectral intensities on the same grid were all within a tolerance of 0.005 of each other. 
    \item Statistical filtering: 
    To avoid potentially unphysical spectra, we filtered out any spectra that are beyond $3\sigma$ of the mean spectrum for each element, which is the same procedure used in Ref.~\citenum{kharel2025omnixas}. 
\end{enumerate}
To speed up the spectral UMAP analysis, the post-processed spectral database was down sampled to about 25\% of its original size, yielding 73,958 spectra. Although the down-sampling changes the exact shape of the UMAP pattern, it does not change our conclusions.

\subsection{XANES Bird}
The resulting UMAP pattern forms a bird-like shape, referred to as the ``XANES Bird," with the head at the top of the projection, followed by the wings, body, and tail moving towards the bottom. It is important to note that the exact shape of the UMAP pattern is highly variable and depends on the choice of parameters. However, we found that the specific clustering of spectra did not change significantly with different parameter choices; only their relative positions and approximate shape and size were affected. We chose a specific set of parameters ($n_{neighbors} = 94$,  $n_{jobs} = 6$, $random\text{ }state = 1579$, and $d_{min} = 0.1$) for ease of visualization. Since UMAP embeddings are rotation invariant, we rotated our UMAP plot 180 degrees purely for visualization purposes.

\begin{figure}[tbh!]
    \centering
    \includegraphics[width=\linewidth]{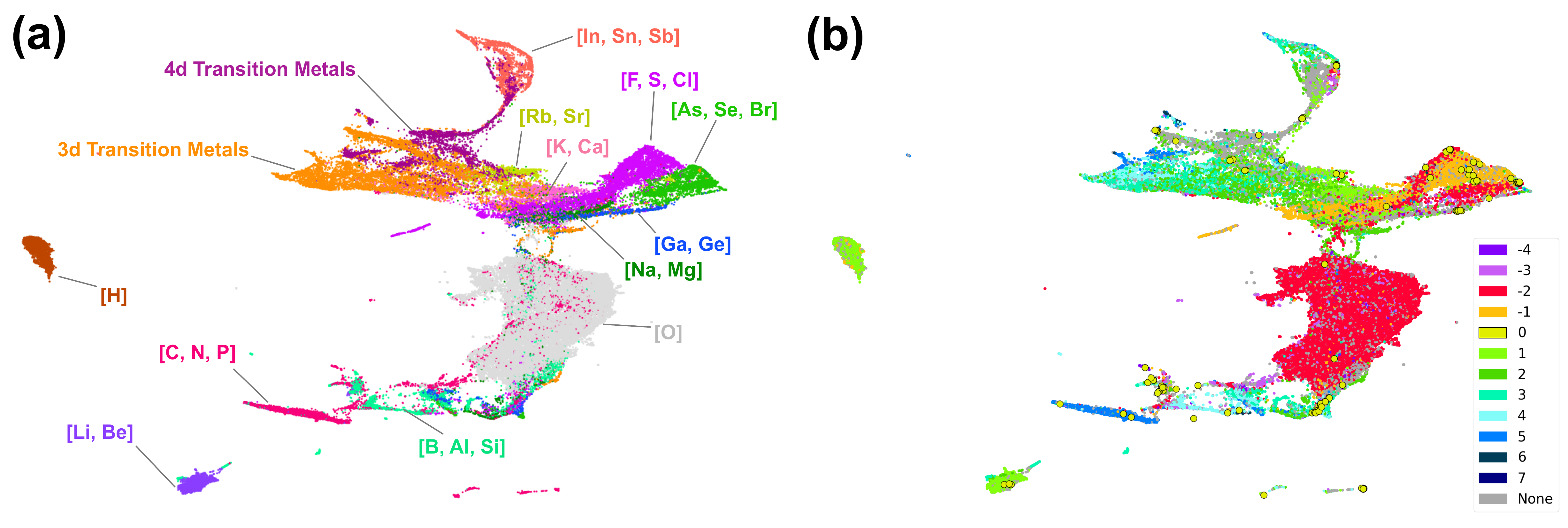}
    \caption{XANES bird: FEFF K-edge XANES spectral UMAP of 47 elements. (a) UMAP colored by clustered sections of elements in the periodic table. (b) UMAP colored by the oxidation state of the absorber atom. ``None" corresponds to cases where BVAnalyzer could not calculate the oxidation state. Zero oxidation state is shown with larger yellow dots outlined in black.}
    \label{fig:UMAPs}
\end{figure}

As shown in Figure~\ref{fig:UMAPs}a, the spectral UMAP is well separated by regions containing subsets of elements with chemical similarities. More specifically, the head of the UMAP bird consists of the three heaviest elements in the analysis: In, Sn, and Sb. 4d transition metals made up a large part of the neck and upper left wing of the bird, with the neck consisting predominantly of Ag. 3d transition metals dominated the lower left wing of the bird and separated well from the region of 4d transition metals. Most alkali and alkaline earth metals, separated by row, cluster predominantly between the two wings of the bird, and many of the halogen and chalcogen elements (F, S, Cl, Se, Br) cluster in the right wing. The body of the bird is overwhelmingly O, and the tail is made up of a mix of well-separated clusters by either main group nonmetals (C, N, P) or main group metals/metalloids (B, Al, Si). For light elements, there are two predominantly isolated clusters, one for H and one for Li and Be. In general, elements progress from heavier to lighter when moving from top to bottom and left to right. Most regions of the periodic table with metallic or semimetallic properties cluster mostly by row, while nonmetallic clustering is less clearly interpretable. The fact that the spectral UMAP clusters are grouped by specific regions of the periodic table provides promise that an ML algorithm would be able to distinguish various sections of the periodic table based on the K-edge XAS spectral shape. 

Figure~\ref{fig:UMAPs}b shows the same UMAP projection but colored by OS computed using pymatgen's BVAnalyzer class \cite{ong2013python}, which provides a convenient way to visualize the OS distribution by element type. Some of the structures analyzed returned ``None" values, including charge zero states. Thus, in the case of single-element systems, we manually assigned an OS of zero to the site. Overall, positive OS values are located on the left side. For negative OS, -1 and -2 concentrate in the right wing and the body, while -3 and -4 OS are more scattered throughout the UMAP. Elements with a single dominating OS exhibit distinct clusters, e.g., O$^{2-}$, S$^{2-}$, Cl$^-$, F$^-$, H$^+$, Li$^+$ and Na$^+$, and most OS regions show considerably clear separation from each other. This suggests that elements with different OS are distinguishable by their XAS spectra. Additionally, we observe that elements with similar OS often (but not always) appear closer to each other on the UMAP. In contrast, elements with multiple possible OS (e.g., 3d and 4d transition metals located in the left wing) form a smooth color gradient in the UMAP pattern, corresponding roughly to +1 to +5 charge states. 

The UMAP pattern reveals machine learnable trends in the latent space of XANES spectra of 47 elements, which is sensitive to not only the element identity, but also important local chemical properties like OS. This strong structure-spectrum-property correlation forms the basis for developing universal spectral models that span the entire periodic table for high-throughput, real-time XANES spectral analysis.


\section{Conclusion}
AI/ML is transforming the field of XAS analysis with successful adoption of a suite of ML tools, such as multilayer perceptrons, graph neural networks, autoencoders, transformers, and diffusion models. In this Perspective, we advocate infrastructure development for an AI-driven XAS analysis pipeline with a focus on benchmarks, workflow software, databases, and AI/ML models. We highlight two important research problems in this AI-driven approach. The first is to reconcile the discrepancies between simulation and experiment. By transforming experimental spectra to a simulation-like representation, the accuracy of the ML models can be significantly improved. In the second problem, we use spectral latent space analysis to reveal common trends in the material spectral database across 47 elements. This analysis underscores the benefit of developing universal XAS models for the whole periodic table. We envision that the transition from traditional methods (such as those based on empirical fingerprints or first-principles simulations) to AI-driven methods can eliminate the entry barrier for non-experts, reduce computational cost, and enable real-time XAS analysis. 

\section{Acknowledgment}
This research is supported by the U.S. Department of Energy, Office of Science, Office of Basic Energy Sciences, Award No. FWP PS-030.
This research used Theory and Computation resources of the Center for Functional Nanomaterials (CFN), which is a U.S. Department of Energy Office of Science User Facility, at Brookhaven National Laboratory under Contract No. DE-SC0012704.

This material is based upon work supported by the U.S. Department of Energy, Office of Science, Office of Advanced Scientific Computing Research, Department of Energy Computational Science Graduate Fellowship under Award Numbers DE-SC0023112
and DE-SC0024386. This report was prepared as an account of work sponsored by an agency of the United States Government. Neither the United States Government nor any agency thereof, nor any of their employees, makes any warranty, express or implied, or assumes any legal liability or responsibility for the accuracy, completeness, or usefulness of any information, apparatus, product, or process disclosed, or represents that its use would not infringe privately owned rights. Reference herein to any specific commercial product, process, or service by trade name, trademark, manufacturer, or otherwise does not necessarily constitute or imply its endorsement, recommendation, or favoring by the United States Government or any agency thereof. The views and opinions of authors expressed herein do not necessarily state or reflect those of the United States Government or any agency thereof.

%

\end{document}